\documentstyle[12pt]{article}
\title{\bf Classical and quantum wormholes in a 5D compact Kaluza-Klein theory }
\author{F. Darabi\thanks{e-mail: f-darabi@azaruniv.edu} 
\\
{\small Department of Physics, Azarbaijan University of Tarbiat
Moallem, 53714-161 Tabriz, Iran.}}
\begin{document}
\maketitle
\vspace{15mm}
\begin{abstract}
We study the classical and quantum Euclidean wormholes for an empty (4+1) dimensional Kaluza-Klein universe with a positive cosmological constant and a spatially flat Robertson-Walker type metric. It is shown that classical wormholes do not exist neither for the spacetime sector nor the extra dimensional
sector of this model, but two spectrums of quantum wormholes as the solutions of Wheeler-DeWitt equation exist which are consistent with the Hawking-Page conjecture about the necessary boundary conditions. In the spectrum where
the external scale factor $R$ shapes the quantum wormholes the internal scale factor $a$ may play the role of effective matter source, and in the spectrum where the internal scale factor $a$ shapes the quantum wormholes the external scale factor $R$ may play the role of effective matter source.
\end{abstract}

\vspace{15mm} 

\newpage

\section{Introduction}

Classical Euclidean wormholes are usually considered as Euclidean metrics
that consist of two asymptotically flat regions connected by a
narrow throat (handle). Euclidean wormholes have been studied mainly as
instantons, namely solutions of the classical Euclidean field
equations. In general, such wormholes can represent quantum
tunneling between different topologies. They are possibly useful
in understanding black hole evaporation \cite{Haw}; in allowing
nonlocal connections that could determine fundamental constants;
and in vanishing the cosmological constant $\Lambda$ \cite{Col1}-\cite{Col3}.
They are even considered as an alternative to the Higgs mechanism
\cite{Mig}. Consequently, such solutions are worth finding.

Unfortunately, they exist for certain special kinds of matter
\cite{Gid1}-\cite{Gid5}. For example, they exist for an imaginary minimally
coupled scalar field \cite{HP}, but do not exist for pure gravity.
Due to limited known classical wormhole solutions, Hawking and page
advocated a different approach in which wormholes were regarded, not
as solutions of the classical Euclidean field equations, but as
solutions of the quantum mechanical Wheeler-DeWitt equation. These wave functions have to obey certain boundary conditions in order that they represent wormholes. The boundary conditions seem to be that: (i) the wave function is 
damped for large 3-geometries, namely 3-dimensional space-like
hypersurfaces with large scale factor; (ii) the wave function is
regular in some suitable way when the tree-geometry collapses to
zero \cite{HP}. The first condition expresses the fact that spacetime should be Euclidean when tree geometries become infinite. The second condition
should reflect the fact that spacetime is nonsingular when tree
geometries degenerates, namely the wave function should not
oscillate an infinite number of time. Therefore, in general, an open and interesting problem is to find new quantum wormholes satisfying these boundary
conditions \cite{QW1}-\cite{QW3}.  

To the authors knowledge, the study of Euclidean wormholes is usually limited
to 4-dimensional spacetime including appropriate forms of matter. Although
we are living in a 4-dimensional universe filled with different matter sources, however, according to Kaluza-Klein theory, our world may be of higher dimensional spacetime without ordinary forms of matter sources. Hence, our strong motivation for studying the 5-dimensional Euclidean wormholes in the spirit of Kaluza-Klein theory is to answer the following equivalent question: ``Are there Euclidean wormholes, shaped by the spacetime dimensions, in an empty higher dimensional universe where the extra dimensions play the role of {\it effective} matter sources?", or ``Are there Euclidean wormholes, shaped by the extra dimensions, in an empty higher dimensional universe where the spacetime dimensions play the role of {\it effective} matter sources?  In looking for such higher dimensional solutions, we consider an empty 5-dimensional Kaluza-Klein cosmology with a cosmological constant and a Robertson-Walker type metric having two dynamical variables, the usual scale factor $R$ and the internal scale factor $a$. First, we investigate the existence of classical Euclidean wormholes. Then, by constructing the corresponding quantum cosmology and Wheeler-DeWitt equation, we look for some solutions of the wave function, so called quantum wormholes, which other
than the previously mentioned boundary conditions can satisfy the boundary conditions: (i) the wave function is damped for large extra dimensional-geometries, (ii) the wave function is regular in some suitable way when the extra dimensional-geometry collapses to zero. 

\section{Classical wormholes}

To begin with, we study the Euclidean form of the metric considered in \cite{D1}-\cite{D4} in which the space-time is assumed to be of Robertson-Walker type
having a (3+1)-dimensional space-time part and an internal space
with dimension $D$. We adopt a real chart $\{t, r^{i}, \rho^{\alpha}\}$
with $t$, $r^{i}$, and $\rho^{\alpha}$ denoting the Euclidean time, space
coordinates and internal space dimensions, respectively. We,
therefore, take
\begin{equation}\label{0}
ds^2=N^2(t) dt^2+R^2(t)\frac{dr^i
dr^i}{(1+\frac{kr^2}{4})^2}+a^2(t) \frac{d\rho^{\alpha}
d\rho^{\alpha}}{(1+k^\prime \rho^2)}, 
\end{equation}
where $N(t)$ is the lapse function, $R(t)$ and $a(t)$ are the
external and internal scale factors, respectively; $r^2 \equiv r^i r^i (i=1, 2, 3), \rho^2\equiv \rho^{\alpha}
\rho^{\alpha} ({\alpha}=1, ... D)$, and $k, k^\prime=0, \pm1$, reflecting flat, open or closed type of four-dimensional universe and
$D$-dimensional space with compact topology $S^D$. In agreement with the present observational constraints we require $k=0$, namely a flat universe. For simplicity, we also assume the internal space to be flat $k^\prime =0$
and $D=1$ dimensional. This flatness assumption is motivated by the possibility of the compact spaces to be flat or hyperbolic in ``{\it accelerating cosmologies from compactification}'' scenarios. Moreover, since the lapse function is an arbitrary function of time because of {\it time reparametrization invariance} property, we take for simplicity $N(t)=1$. Therefore, we work with the metric \begin{equation}\label{1}
d s^2 = {dt}^2+R^2 (t) {dr^i \: dr^i}+a^2 (t)
{d\rho}^2.
\end{equation}
The curvature scalar corresponding to the metric (\ref{1})
is obtained as
\begin{equation}\label{2}
{\cal R}=-6 \left[\frac{\ddot{R}}{R}+\frac{{\dot{R}}^2}{R^2}\right] -2\left(\frac{
\ddot{a}}{a}+ 3 \frac{\dot{R}}{R} \frac{\dot{a}}{a}\right),
\end{equation}
where a dot represents differentiation with respect to $t$. Substituting
this result into Einstein-Hilbert action with a cosmological constant
$\Lambda$, namely  
\begin{equation}\label{3}
I=\int\!\sqrt{-g} ({\cal R}-\Lambda) dt\:d^3 r\:d\rho,
\end{equation}
and integrating over spatial dimensions gives an effective Lagrangian $L$ in the mini-superspace ($R$,$a$) as 
\begin{equation}\label{4}
L=-\frac{1}{2} R a {\dot{R}}^2-\frac{1}{2}R^2 \dot{R}\dot{a}+\frac{1}{6} \Lambda R^3 a.
\end{equation}
By changing the variables as 
\begin{equation}\label{5}
X= \ln R,\:\:\: Y=\ln a
\end{equation}
$L$ takes on the form
\begin{equation}\label{6}
L=\frac{1}{2}e^{3X+Y}[-\dot{X}^2-\dot{X}\dot{Y}+\frac{1}{3}\Lambda].
\end{equation}
A further change of variables as
\begin{equation}\label{7}
x=\frac{1}{4}(3X+Y), \:\:\: y=\frac{1}{4}(X-Y),
\end{equation}
leads us to the following Lagrangian
\begin{equation}\label{8}
L=e^{4x}[-\dot{x}^2+\dot{y}^2+\frac{1}{6}\Lambda].
\end{equation}
A Legendre transformation on (\ref{8}) provides us with the corresponding Hamiltonian
\begin{equation}\label{9}
H=\frac{1}{4}e^{-4x}[-p_{x}^2+p_{y}^2-\frac{2}{3}\Lambda e^{8x}].
\end{equation}
Due to the {\it time reparametrization invariance} property of the
present cosmological model the zero energy condition $H=0$ is imposed\footnote{We know that general relativity is a {\it time reparametrization invariant}
theory. Every theory which is invariant under an action casts into the constraint
systems. Therefore, general relativity is a constraint system whose constraint
is the zero energy condition $H=0$ \cite{Dirac,ADM}.}
\begin{equation}\label{10}
[p_{x}^2-p_{y}^2+\frac{2}{3}\Lambda e^{8x}]=0.
\end{equation}
The Euclidean solutions in terms of $R(t)$ and $a(t)$ subject to the
Hamiltonian constraint (\ref{10}) are obtained
\begin{equation}\label{11}
R(t)=\frac{\sqrt{-6\sqrt{6\Lambda}[C_1 \sin(\frac{\sqrt{6\Lambda}}{3}t)-C_2 \cos(\frac{\sqrt{6\Lambda}}{3}t)]}}{\sqrt{6\Lambda}},
\end{equation}
\begin{equation}\label{12}
a(t)=C_3\exp\left[-\int\left(\frac{\dot{R}}{R}+\frac{\Lambda}{3}\frac{R}{\dot{R}}\right) dt\right].
\end{equation}
Demanding for real solutions of $R(t)$ and $a(t)$ requires the cosmological constant to be positive. It is obvious that in this case, the oscillating
solution (\ref{11}) does not support a classical Euclidean wormhole which is usually considered as Euclidean metric that consists of two {\it asymptotically flat} regions connected by a narrow throat $R_0$. Therefore, no classical Euclidean wormhole exists for this model with positive cosmological constant and the internal scale factor $a(t)$ can not play the role of an effective matter source suitable for classical Euclidean wormholes. Note that, 
if we consider the internal scale factor, regardless of the compactification, in the same foot as the external one, in principle we may then look for the classical Euclidean wormhole whose configuration is shaped by the internal scale factor $a$. However, if such a consideration was even made, the solution (\ref{12}) would not support a classical Euclidean wormhole because it could not represent two {\it asymptotically flat} regions connected by a narrow throat $a_0$.

In fact, it is well known that classical Euclidean wormholes can occur if the Ricci tensor has negative eigenvalues somewhere on the manifold \cite{Chee}. This is necessary but not sufficient condition for their existence
and is related to the implication of a theorem of Cheeger and Glommol.
The energy-momentum tensors of an axion field and of a conformal
scalar field are such that, when coupled to gravity, the Ricci
tensor has negative eigenvalues. However, for example, in the case of minimally coupled real scalar field the Ricci tensor has not negative eigenvalues and a complex scalar field is necessary for Euclidean wormholes to occur as follows. 

Let us consider the action of an ordinary scalar field minimally coupled to gravity\footnote{We use the units in which $\frac{8\pi G}{3}=1$.}
\begin{equation}\label{1'}
{\cal S}=\int d^4x \sqrt{-g}\left[{\cal R}-\frac{1}{2}\nabla_{\mu}\phi \nabla^{\mu}\phi-V(\phi)\right],
\end{equation}
where $V(\phi)$ is the potential of the scalar field. 
The energy momentum tensor of the field is derived by
varying the action in terms of $g^{\mu \nu}$
\begin{eqnarray}\label{2'}
T_{\mu \nu}&=&-\frac{2}{\sqrt{-g}}\frac{\delta{\cal S}}{\delta g^{\mu \nu}}\\ \nonumber
&=&\nabla_{\mu}\phi \nabla_{\nu}\phi-g_{\mu
\nu}\left[\frac{1}{2}\nabla_{\lambda}\phi\nabla^{\lambda}\phi+V(\phi)\right].
\end{eqnarray}
The trace is also obtained as 
\begin{equation}\label{3'}
T=-\nabla_{\mu}\phi \nabla^{\mu}\phi-4V(\phi).
\end{equation}
The Einstein equations 
\begin{equation}\label{4'}
R_{\mu \nu}=T_{\mu \nu}-\frac{1}{2}g_{\mu \nu}T,
\end{equation}
give
\begin{equation}\label{5'}
R_{\mu \nu}=\nabla_{\mu}\phi \nabla_{\nu}\phi+ g_{\mu \nu}V(\phi),
\end{equation}
which shows $R_{\mu \nu}$ can never be negative in Euclidean space $g_{\mu \nu}=(++++)$ for $V(\phi)\geq 0$ unless we consider
a purely imaginary scalar field, i.e., let $\phi \rightarrow i\phi$ \cite{Coule}.

In the same way, we may study the necessary condition for classical Euclidean wormhole in the present model. The full (4+1)-dimensional Einstein equations are obtained by varying the action (\ref{3}) with respect to the metric $g_{AB}$ as
\begin{equation}
R_{AB}=\Lambda g_{AB},
\end{equation}
where $R_{AB}$ is the (4+1)-dimensional Ricci tensor. It is then obvious that the Ricci tensor has never negative eigenvalues for positive cosmological constant and the necessary condition for the existence of wormhole solution fails. This confirms the previous result that no classical Euclidean wormhole exists for this model
with positive cosmological constant.

\section{Quantum wormholes}

To obtain the Wheeler-DeWitt equation we may use the following quantum mechanical replacements
\begin{equation}\label{13}
p_x \rightarrow -i\frac{\partial}{\partial x}\:\:\:\:,\:\:\:\: p_y
\rightarrow -i\frac{\partial}{\partial y},
\end{equation}
by which the Hamiltonian constraint (\ref{10}) is transformed to the 
Wheeler-DeWitt equation 
\begin{equation}\label{14}
\left[\frac{\partial^2}{\partial
x^2}-\frac{\partial^2}{\partial
y^2}+\frac{2}{3}\Lambda e^{8x}\right]\Psi(x, y)=0, 
\end{equation}
where $\Psi(x, y)$ is the wave function of the universe in the
$(x, y)$ mini-superspace. The wave function is easily separable as 
$\Psi(x, y)=\psi(x) \phi(y)$. Substituting this into the Wheeler-DeWitt equation
yields
\begin{equation}\label{15}
\left[\frac{d^2}{d
x^2}+\frac{2}{3}\Lambda e^{8x}-B^2\right]\psi(x)=0, 
\end{equation}
\begin{equation}\label{16}
\left[\frac{d^2}{d
y^2}-B^2\right]\phi(y)=0, 
\end{equation}
where $B$ is a constant. These differential equations are easily solved.
Let us define 
\begin{equation} \label{17}
\left\{ \begin{array}{ll} \sigma^2=e^{8x}, \\
\\
m^2=\frac{B^2}{16},\\
\\
n^2=\frac{\Lambda}{16}.
\end{array}
\right.
\end{equation}
Then, Eq.(\ref{15}) is converted to the Bessel's differential equation
\begin{equation}\label{18}
\sigma\frac{d}{d\sigma}\left(\sigma\frac{d\psi}{d\sigma}\right)+(n^2\sigma^2-m^2)\psi=0, \end{equation}
whose solution is known as Bessel or Neumann function:
$$
\psi(\sigma)=J_{B/4}\left(\frac{\sqrt{\Lambda}}{4}\sigma\right),
$$
$$
\psi(\sigma)=N_{B/4}\left(\frac{\sqrt{\Lambda}}{4}\sigma\right).
$$
On the other hand, Eq.(\ref{16}) has simple solution  
$$
\phi(y)=Ce^{\pm By},
$$
where $C$ is a constant. By transforming back to the original variables $R$ and $a$, we obtain 
\begin{equation}\label{19}
\Psi(R, a)=C\left(\frac{R}{a}\right)^{\pm{B}/{4}}\left[\alpha J_{{B}/{4}}\left(\frac{\sqrt{\Lambda}}{4}R^3a\right)+\beta
N_{{B}/{4}}\left(\frac{\sqrt{\Lambda}}{4}R^3a\right)\right]. 
\end{equation}
We now impose the boundary conditions of quantum wormhole solutions. First,
we notice that we have two dynamical variables $R$ and $a$ together with a parameter $\Lambda$ in the argument of the Bessel function. Therefore, we may consider the following three cases:

\subsection*{Case 1: Variable $R$ for given $a, \Lambda$}

In order for the wave function not to be diverged at large scale factor $R$, the positive sign in the power of the term $(R/a)$ in (\ref{19}) should be removed. The wormhole boundary condition at $R\rightarrow 0$ is also satisfied just for the Bessel function of the $J$ kind. Therefore, we have
\begin{eqnarray}\label{20}
\Psi(R\rightarrow \infty, a, \Lambda)&=&\alpha C\left(\frac{R}{a}\right)^{-{B}/{4}}\frac{4}{\sqrt{\Lambda}R^3a}\sin\left(\frac{\sqrt{\Lambda}}{4}R^3a-\frac{B\pi}{8}\right)\\
\nonumber
&=&\frac{4\alpha C}{\sqrt{\Lambda}}R^{-(3+B/4)}a^{(B/4-1)}\sin\left(\frac{\sqrt{\Lambda}}{4}R^3a-\frac{B\pi}{8}\right),
\end{eqnarray}
where use has been of the asymptotic behaviour of the Bessel function of the $J$ kind. This shows the wave function satisfies the boundary condition in that decays to zero at large scale factor $R$ provided $B>-12$. On the other hand, using the behaviour of
the Bessel function in the limit $R\rightarrow 0$ we obtain
\begin{equation}\label{21}
\Psi(R\rightarrow 0, a, \Lambda)=\alpha C\frac{(\sqrt{\Lambda}/4)^{B/4}(Ra)^{B/2}}{1.3.5
... (B/2 +1)},
\end{equation}
which shows the wave function satisfies the other boundary condition in that it is regular in some suitable way when the scale factor collapses to
zero. Unlike the classical solutions, the quantum solutions with variable $R$ admit quantum wormholes, where the internal scale factor and cosmological
constant may play the role of an effective matter source. 

\subsection*{Case 2: Variable $a$ for given $R, \Lambda$}

In order for the Kaluza-Klein theory to be in agreement with current experimental
observations of the observable Lorentzian universe, it is a well known fact that the internal space should be a compact space having a very small internal scale factor, namely $a$. In the present work, we follow the Kaluza-Klein idea by assuming a cylinderical condition on the extra dimension. Note however that, if one assumes (as we did) this scale factor to be dependent on the {\it Euclidean} time parameter $t$, one may no longer guarantee the internal scale factor to be small in terms of the Euclidean time. In fact, the non-observability
of extra dimension concerns just about the observable Lorentzian universe, so the question of observability (largeness) or non-observability (smallness)
of extra dimension in the non-observable Euclidean universe is meaningless. It is easily seen that the solution (\ref{12}) is an oscillating one in terms of Euclidean time. Therefore, the internal scale factor is not fixed to an small size, rather it oscillates between some small and large sizes in terms of Euclidean time. 

Even, if somehow the internal space would be kept small {\it classically}, there would be no reason from quantum cosmological considerations to forbid the scale factor $a$ being variable taking large as well as small
values. This is because the classical time parameter $t$ has no special role in quantum cosmology and in order for the boundary conditions are imposed on the wavefunction, the scale factor $a$ as a collective coordinate (irrespective of time parameter $t$) in the two dimensional ($R, a$) mini-superspace is allowed to take large as well as small values, similar to the scale factor $R$. Actually, taking large values for the scale factor $a$ (as a coordinate in the mini-superspace) in imposing the boundary conditions does not necessarily mean that the scale factor $a$ becomes classically large. This is similar to the problem of a particle in a one dimensional potential barrier, where the particle is classically confined within the distance $d$ having positive kinetic energy and is not allowed to take positions
$x>a$ where the kinetic energy is negative, but, quantum mechanically the coordinate $x$ in the wavefunction $\psi(x)$ is allowed to take large values
$x>a$. Based on the above discussion, we consider the cases where the scale factor $a$ as a variable approaches to the extremely small and large values. 

For given values of $R, \Lambda$, the wave function does not diverge at large $a$, provided that the negative sign in the power of the term $(R/a)$ in (\ref{19}) is removed. Therefore, we have
\begin{eqnarray}\label{22}
\Psi(a\rightarrow \infty, R, \Lambda)&=&\alpha C\left(\frac{R}{a}\right)^{{B}/{4}}\frac{4}{\sqrt{\Lambda}R^3a}\sin\left(\frac{\sqrt{\Lambda}}{4}R^3a-\frac{B\pi}{8}\right)\\
\nonumber
&=&\frac{4\alpha C}{\sqrt{\Lambda}}R^{(B/4-3)}a^{-(B/4+1)}\sin\left(\frac{\sqrt{\Lambda}}{4}R^3a-\frac{B\pi}{8}\right),
\end{eqnarray}
This shows the wave function decays to zero at large $a$ provided $B>-4$. On the other hand, in the limit $a\rightarrow 0$ we obtain
\begin{equation}\label{23}
\Psi(a\rightarrow 0, R, \Lambda)=\alpha C\frac{(\sqrt{\Lambda}/4)^{B/4}R^{B}}{1.3.5
... (B/2 +1)}.
\end{equation}
This shows the wave function is regular when the radius $a$ collapses to
zero. Altogether, we realize that the case of variable $a$ represents a quantum wormhole corresponding to the internal scale factor $a$, where the external scale factor and cosmological constant may play the role of an effective matter source.

\subsection*{Case 3: Variable $\Lambda$ for given $R, a$}

In the model Universe considered here, the cosmological constant $\Lambda$
is just a parameter. However, in the Multiverse models (Universes with different physical constants) $\Lambda$ is considered as a variable \cite{Tegm,Davis}. It is obvious that the quantum wormhole boundary conditions, which are
imposed on the scale factors $R, a$, are not relevant in this case because
$\Lambda$ is not a scale factor. Therefore,
one may study the probabilistic interpretation
of the wavefunction (\ref{19}) assuming a variable $\Lambda$ over a Multiverse. For given values of $R, a$, with $B\neq 0$ we have
\begin{eqnarray}\label{24}
\Psi(\Lambda\rightarrow \infty, R, a)&=&\alpha C\left(\frac{R}{a}\right)^{\pm{B}/{4}}\frac{4}{\sqrt{\Lambda}R^3a}\sin\left(\frac{\sqrt{\Lambda}}{4}R^3a-\frac{B\pi}{8}\right),
\end{eqnarray}
which shows the wave function is vanishing for large cosmological constant $\Lambda$. In the limit of zero cosmological constant we have
\begin{equation}\label{25}
\Psi(\Lambda\rightarrow 0, R, a)=\alpha C\frac{(\sqrt{\Lambda}/4)^{B/4}R^{B}}{1.3.5
... (B/2 +1)}\rightarrow 0.
\end{equation}
This again shows the wave function is vanishing for zero cosmological constant $\Lambda=0$.

\section*{Conclusion}

Euclidean wormholes are of particular importance from
macroscopic and microscopic points of view, in quantum gravity. 
In particular,  microscopic wormholes are studied as instantons, namely the saddle points in the Euclidean path integrals. Thus, one can use them to give a semi-classical treatment in the dilute wormhole approximation where
the interaction between the large scale ends of wormholes is
neglected. Moreover, any new wormhole solution may provide a new contribution for black hole evaporation in theories with reasonable matter content, and it is supposed to play an important role in vanishing the cosmological constant. 
In a generalization to higher dimensional models, we have studied the classical and quantum wormhole solutions for a typical empty (4+1) dimensional Kaluza-Klein universe with a positive cosmological constant and a Robertson-Walker type metric. We were originally motivated to investigate whether higher dimensions in an empty universe can play the role of matter source to produce the Euclidean
wormholes. However, the motivation for the study of extra dimension $a$ as effective matter source for the Euclidean wormholes in a $R$ variable universe led us to the interesting result: We have shown that neither classical wormholes for variable $R$ nor classical wormholes for variable $a$ exist for this model, but two spectrums of quantum wormholes have been obtained, in one spectrum the external scale factor $R$ shapes the quantum wormholes while the internal scale factor $a$ may play the role of effective matter source,
and in the other spectrum the internal scale factor $a$ shapes the quantum wormholes while the external scale factor $R$ may play the role of effective matter source.

Assuming the cosmological constant $\Lambda$ as a variable in a Multiverse scenario, we have examined the behaviour of the wavefunction for $\Lambda\rightarrow \infty$ and $\Lambda\rightarrow 0$. We have shown that the wavefunction is vanishing for both limits and this is interpreted, in the context of {\it Anthropic principle} \cite{Barrow}, in favor of universes with {\it finite-nonzero} cosmological constants.

\section*{Acknowledgment}

This research was supported by a grant/research fund Number 217/D/2660 from Azarbaijan University of Tarbiat Moallem.

\end{document}